\documentclass[aps]{revtex4}  

\usepackage{graphicx}                                                %

\begin{document}
\title{Recovery of continuous wave squeezing at low
frequencies}

\author{W.~P.~Bowen, R.~Schnabel, N.~Treps, H.-A.~Bachor, and
P.~K.~Lam} \affiliation{Department of Physics, Faculty of Science, \\
The Australian National University, A.C.T. 0200, Australia.}
\email{warwick.bowen@anu.edu.au}

\begin{abstract} We propose and demonstrate a system that produces
squeezed vacuum using a pair of optical parametric amplifiers.  This
scheme allows the production of phase sidebands on the squeezed vacuum
which facilitate phase locking in downstream applications.  We observe
strong, stably locked, continuous wave vacuum squeezing at frequencies
as low as 220~kHz.  We propose an alternative resonator configuration
to overcome low frequency squeezing degradation caused by the optical
parametric amplifiers.
\end{abstract}

\maketitle   

Squeezed states of light have the potential to improve the sensitivity
of interferometric \cite{Xiao87}, spatial \cite{Fabre99} and
spectroscopic \cite{Polzik92} measurements.  In almost all experiments
to date the squeezing has been observed at frequencies above 1~MHz, at
lower frequencies most laser sources are classically noisy.  Many
interesting spatial and interferometric signals occur in the Hz to kHz
regime \cite{Boccara80,Meshkov00}, to study these dynamics it is
useful to produce squeezing at lower frequencies.  Two experiments
using pulsed laser light have produced squeezing in the kHz regime
\cite{Bergman91,Hirano92}.  However, for applications that involve
resonators or have peak power limitations (e.g. optical damage) such
as advanced gravitational wave detection \cite{Meshkov00}, the
squeezed light should be continuous wave.

In this paper we report the demonstration of continuous wave squeezing
in the kHz regime.  By operating a pair of optical parametric
amplifiers (OPAs) within a Mach-Zehnder interferometer we produced a
squeezed vacuum at frequencies as low as 220 kHz.  In this
configuration it was possible to produce bright sidebands on the
squeezed vacuum.  These sidebands acted as a phase reference that
allowed the phase of the vacuum to be locked in a downstream homodyne
detector.  The frequency of the squeezing reported here was limited
from below by noise introduced inside the OPAs.  We propose an
extension to this work that should eliminate these techincal issues
and produce squeezing at much lower frequencies.

\begin{figure}[ht]
  \begin{center}
  \includegraphics{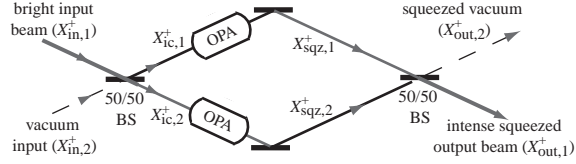}
  \end{center}
  \caption{Squeezed vacuum production via classical noise cancellation
  using two OPAs.}
  \label{TheoryPic}
\end{figure}
Noise cancellation schemes have been proposed to enhance squeezing
utilizing Kerr non-linearity in fibers \cite{Shirasaki90} and second
harmonic generation \cite{Ralph95}.  The scheme presented here is
similar to that of \cite{Shirasaki90} but uses optical parametric
processes.  A single laser beam is split on a 50/50 beam splitter and
used to seed a pair of identical OPAs, each consisting of a
$\chi^{(2)}$ non-linear medium inside an optical resonator.  The OPA
output fields are then re-combined in phase on another 50/50 beam
splitter as shown in Fig.  \ref{TheoryPic}.  The phase of the seed
field relative to a non-resonant second harmonic pump field dictates
the seed amplification of each OPA. This phase is controlled to
de-amplify the seed, resulting in amplitude squeezed output fields.
In this configuration the overall non-linearity $\Upsilon$ of each
resonator, which is dependant on the crystal non-linearity and pump
power, becomes real and negative.  A treatment of optical parametric
oscillation which uses the linearized formalism of quantum mechanics
and the mean field approximation is given in \cite{Lam99}.  Extending
this to optical parametric amplification we obtain the fourier domain
amplitude quadrature operators $X^{+}_{{\rm sqz} \!  , j}$ for the
fields exiting the OPA output couplers
\begin{eqnarray}
\label{OPAoutput}
X^{+}_{{\rm sqz} \!  , j} &=& \frac{\sqrt{2 \gamma_{{\rm oc}}}}{i
\omega \!  - \!  \Upsilon \!  + \!  \gamma} \!  \left (\sqrt{2
\gamma_{{\rm ic}}} X^{+}_{{\rm ic} \!  , j} \!  + \!  \sqrt{2
\gamma_{l}} X^{+}_{l \!  , j} \right ) \nonumber \\
&+& X^{+}_{{\rm oc} \!  , j} \!  \left (\frac{2 \gamma_{{\rm oc}}}{i
\omega \!  - \!  \Upsilon \!  + \!  \gamma} \!  - \!  1 \right )
\end{eqnarray}
$X^{+}_{{\rm ic} \!  , j}$ and $X^{+}_{{\rm oc} \!  , j}$ are the
amplitude quadrature operators of the external fields incident on the
input (ic) and output (oc) couplers of the OPAs; and the vacuum fields
$X^{+}_{l \!  , j}$ result from loss inside each OPA. The two
interacting modes considered here are distinguished by the subscripts
$j \in \{ 1,2 \} $ throughout the paper (see Fig.  \ref{TheoryPic}).
$\gamma_{{\rm ic}}$, $\gamma_{{\rm oc}}$ and $\gamma_{l}$ are the
decay rates respectively due to the input and output coupler
transmitivity, and the intracavity loss; and $\gamma \!= \!
\gamma_{{\rm ic}}\!+\!\gamma_{{\rm oc}}\!+\!\gamma_{l}$ is the overall
resonator decay rate.

The amplitude quadrature operators for the output fields of a 50/50
beam splitter with one intense $X^{+}_{in \!  ,1}$ and one vacuum
$X^{+}_{{\rm in} \!  ,2}$ input field can be expressed as $X^{+}_{{\rm
ic} \!  ,1} \!=\!  (X^{+}_{{\rm in} \!  ,1} \!+\!  X^{+}_{{\rm in}
\!  ,2})/\sqrt{2}$ and $X^{+}_{{\rm ic} \!  ,2} \!=\!  (X^{+}_{{\rm
in} \!  ,1} \!-\!  X^{+}_{{\rm in} \!  ,2})/\sqrt{2}$.  Substituting
these expressions as the seed fields in Eqs.  (\ref{OPAoutput}) we
obtain
\begin{eqnarray}
\label{out2}
X^{+}_{{\rm sqz} \!  , j} \!  &=& \!  \frac{\sqrt{2 \gamma_{{\rm
oc}}}}{i \omega \!  - \!  \Upsilon \!  + \!  \gamma} \!  \left
(\sqrt{\gamma_{{\rm ic}}} \left (X^{+}_{{\rm in} \!  ,1} \!-\!
({\textrm{-}} 1)^{j} X^{+}_{{\rm in} \!  ,2} \right ) \!  + \!
\sqrt{2 \gamma_{l}} X^{+}_{l \!  , j} \right ) \nonumber
\\
\!  &+& \!  X^{+}_{{\rm oc} \!  , j} \!  \left (\frac{2 \gamma_{{\rm
oc}}}{i \omega \!  - \!  \Upsilon \!  + \!  \gamma} \!  - \!  1 \right
)
\end{eqnarray}
The interference of these beams in phase on a 50/50 beam splitter
yields an intense squeezed beam, and a squeezed vacuum, which have
amplitude quadrature operators $X^{+}_{{\rm out} \!  , 1}$ and
$X^{+}_{{\rm out} \!  , 2}$ respectively, given by
\begin{eqnarray}
\label{finalout2}
X^{+}_{{\rm out} \!  , j} \!  &=& \!  \frac{\sqrt{\gamma_{{\rm
oc}}}}{i \omega \!  - \!  \Upsilon \!  + \!  \gamma} \!  \left ( \!  2
\sqrt{\gamma_{{\rm ic}}} X^{+}_{{\rm in} \!  , j} \!  + \!  \sqrt{2
\gamma_{l}} (X^{+}_{l \!  ,1} \!  - \!  ({\textrm{-}} 1
)^{j} X^{+}_{l \!  ,2}) \right ) \nonumber \\
\!  &+& \!  \frac{1}{\sqrt{2}} \!  \left (\frac{2 \gamma_{{\rm oc}}}{i
\omega \!  - \!  \Upsilon \!  + \!  \gamma} \!  - \!  1 \right ) \!
\left (X^{+}_{{\rm oc} \!  ,1} \!  - \!  ( {\textrm{-}} 1)^{j}
X^{+}_{{\rm oc} \!  ,2} \right )
\end{eqnarray}
Note that interfering the beams with a $\pi/2$ phase shift would have
produced an Einstein-Podolsky-Rosen entangled pair \cite{Ou92}.  The
amplitude noise variance spectra ${\bf V}^{+}_{{\rm out} \!  , j}$ of
the output beams can be calculated, $\,{\bf V}=\langle {X}^2 \rangle
\!-\!  {\langle X \rangle} ^2\,$.  Seeding each OPA through its input
coupler and assuming that the fields entering through the output
coupler and through loss are uncorrelated vacuum we obtain
\begin{eqnarray}
\label{Vfinalout2}
{\bf V}^{+}_{{\rm out} \!  , j} &=& \frac{4 \gamma_{{\rm oc}}
(\gamma_{{\rm ic}} {\bf V}^{+}_{{\rm in} \!  , j} \!  + \!
\gamma_{l}) \!  + \!  \omega^{2} \!  + \!  (2 \gamma_{{\rm oc}}
\!  + \!  \Upsilon \!  - \!  \gamma)^{2}}{\omega^{2} \!  + \!
(\Upsilon \!  - \!  \gamma)^{2}}
\end{eqnarray}
These spectral noise distributions are identical to those that would
be produced from two independant OPAs, one seeded with the intense
input beam ($X^{+}_{{\rm in} \!  ,1}$), and one with the vacuum field
entering at the first beam splitter ($X^{+}_{{\rm in} \!  ,2}$) (c.f.
Eqs.~(\ref{OPAoutput})).  Unlike the case for a single vacuum seeded
OPA \cite{Breitenbach95} however, the resonance frequency of each OPA
can be conveniently locked to the frequency of the seed beam and the
vacuum squeezed output can be produced with bright sidebands far
outside the frequency range for squeezing.  These sidebands are
produced by applying anti-correlated phase modulation to the
intracavity fields of the OPAs, and allow the quadrature of the
squeezed vacuum to be locked in downstream applications.

In theory none of the laser noise is carried through onto ${\bf
V}^{+}_{{\rm out} \!  ,2}$.  In reality however, the two OPAs will not
be identical and this will cause classical noise from the intense
input beam to couple into the vacuum output.  By varying the input
beam splitter ratio it is possible to completely compensate for these
differences over some frequency range.  For example balancing the
intensity of the seed beams such that the intensity of the (ideally)
squeezed vacuum output of the system ($X^{+}_{{\rm out} \!  ,2}$) is
minimized suppresses the classical noise from the seed beam
($X^{+}_{{\rm in} \!  ,1}$) at $\omega \!  = \!  0$.  Since the
denominators in eqs.  (\ref{finalout2}) are slowly varying with
$\omega$ while $\omega \!  \ll \!  \gamma \!  - \!  \Upsilon$ the
classical noise is also suppressed throughout this frequency range.
The suppression is only limited by inefficiencies in the mode-matching
between the OPA output beams.

\begin{figure}[ht]
  \begin{center}
  \includegraphics{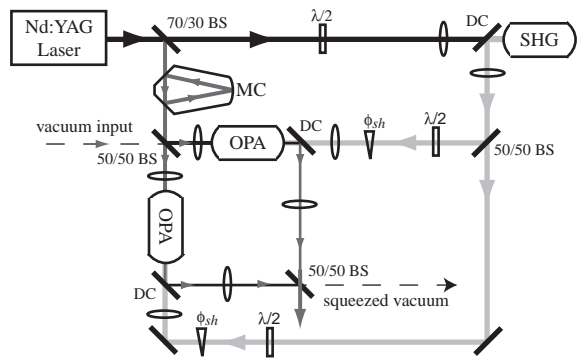}
  \end{center}
  \caption{Schematic diagram of the experiment.  BS: beam splitter,
  DC: dichroic, MC: mode-cleaner, $\lambda/2$: half-wave plate
  $\phi_{{\rm sh}}$ actively controlled phase shift on second
  harmonic beam}
  \label{expt}
\end{figure}
The experimental setup used to generate a locked squeezed vacuum is
shown in Fig.~\ref{expt}.  Two amplitude squeezed beams were produced
in a pair of spatially separated optical parametric amplifiers (OPAs).
The OPAs were optical resonators constructed from hemilithic
MgO:LiNbO$_{3}$ crystals and output couplers.  The reflectivities of
the output couplers were 96\% and 6\% for the fundamental (1064~nm)
and the second harmonic (532~nm) laser modes, respectively.  The OPAs
were pumped with single-mode 532~nm light generated by a 1.5~W Nd:YAG
non-planar ring laser and frequency doubled in a second harmonic
generator (SHG) \cite{Shaddock00}.  Each OPA was seeded with 1064~nm
light after spectral filtering in a mode-cleaner.  An intracavity phase
modulation on this seed enabled control of the length of the OPAs.
The coherent amplitude of the output of each OPA experienced an
amplification dependent on the phase difference between pump and seed
($\phi_{{\rm sh}}$).  A phase modulation on the second harmonic pump
generated an error signal that was used to lock to maximum
deamplification.  In this regime each OPA produced an amplitude
squeezed beam.  The seed power to the OPAs was adjusted so that these
two beams were of equal intensity.
We combined both beams on a 50/50 beam splitter and observed a
visibility between them of 98.6~\%.  One output of this beam splitter
was used to lock the relative phase of the input beams producing
either an intense (1.2~mW), or a weak (0.9~$\mu$W) squeezed beam at
the other output.  This beam was analyzed in a homodyne detector that
was locked so that the phase modulation from the OPAs was not
observed.  Since the OPA outputs were amplitude squeezed, this setup
naturally provided a measurement of the maximally squeezed quadrature
independant of the relative phase of the squeezed beams incident on
the 50/50 beam splitter.  A homodyne visibility of 96~\% was observed
for the intense squeezed beam.  The majority of the power of the weak
squeezed beam came from the residual unmode-matched part of the input
beams and had a non-Gaussian transverse modeshape, consequently the
homodyne visibility observed for this beam was only 28~\%.  The
squeezed vacuum however, arose from the mode-matched part of this beam
and therefore had the same homodyne visibility as the intense squeezed
beam.  The low visibility of the residual power caused the classical
noise that it carried to be poorly detected.  This effect and the
50/50 beam splitter visibilty together led to a predicted optical
supression of approximately 29~dB. The classical noise from the
homodyne local oscillator was electronically suppressed by 64~dB and
was insignificant for our measurements.

\begin{figure}[ht]
  \begin{center}
  \includegraphics{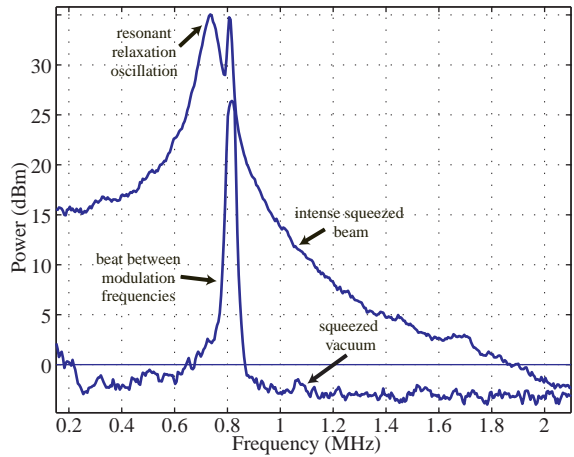}
  \end{center}
  \caption{Squeezing observed on the intense squeezed output (upper
  trace) and on the squeezed vacuum output (lower trace).}
  \label{results}
\end{figure}
\begin{figure}[ht]
  \begin{center}
  \includegraphics{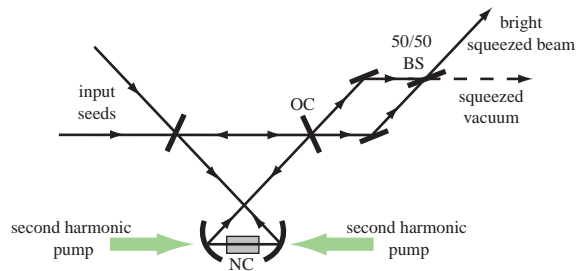}
  \end{center}
  \caption{Proposed scheme to cancel classical noise.  OC: output
  coupler, BS: beam splitter, NC: non-linear crystal.}
  \label{future}
\end{figure}

Fig.~\ref{results} shows the observed squeezing traces for intense
and vacuum squeezed beams.  Both the OPA resonators and the homodyne
detector were phase locked.  The intense squeezed beam was degraded by
the resonant relaxation oscillation of the laser and squeezing was
only observed above 1.9~MHz.  The squeezed vacuum however was observed
to be squeezed from 220~kHz to the end of our measurement range at
2.1~MHz, excluding frequencies between 640~kHz and 870~kHz where our
optical suppression was not high enough to suppress the laser
relaxation oscillation.  The signal at 810~kHz was the beat between
the SHG and OPA locking frequencies.
The low frequency degradation observed in our squeezed vacuum was
caused by uncorrelated acoustic and locking noise introduced within
the OPAs.  Further improvement to the squeezing spectra could only be
achieved by either reducing this noise, or by classically correlating
it.  Employing a ring OPA as shown in Fig.~\ref{future} accomplishes
the second of these alternatives.  A squeezed output is produced by
each of the two directional modes in the OPA. In this configuration
the noise introduced in the resonator is common mode and cancels on
the vacuum output of a 50/50 beam splitter.

The noise cancellation techniques described in this paper is
compatible with existing techniques such as spectrally filtering
the OPA seeds in a high finesse mode-cleaner, or using a servo
controlled laser intensity noise eater.

\label{concl}
In conclusion, we have produced a locked continuous wave squeezed
vacuum with a pair of OPAs.  By phase modulating the OPAs we produced
phase sidebands that act as a phase reference allowing the quadrature
of the squeezing to be locked in downstream applications.  We used
these sidebands to lock a homodyne detector and measured stably locked
vacuum squeezing down to 220~kHz.  We propose an alternative OPA
configuration that should allow the production of vacuum squeezing at
significantly lower frequencies.

The authors acknowledge the Alexander von Humboldt foundation for
support of R.~Schnabel; the Australian Research Council for
financial support; and Dr.~D.~Shaddock for insightful discussion.

\end{document}